\documentclass[a4paper,12pt,oneside]{article}
\usepackage[warn]{mathtext}
\usepackage{graphics}
\usepackage{amsmath}
\usepackage[T2A]{fontenc}
\usepackage{latexsym}
\usepackage[utf8x]{inputenc} 
\usepackage[english]{babel}
\usepackage{graphics}

\title{Half-and-half HFCG with double-end initiation}
\author{S.~Anishchenko$^1$, P.~Bogdanovich$^1$, A.~Gurinovich$^1$, D.~Leonenko$^2$\\\small $^1$Research Institute for Nuclear Problems of BSU, Bobruiskaya str., 11, 220030, Minsk, Belarus\\\small $^2$Electrophysical Laboratory, Smolenskaya str., 15A, 220088, Minsk, Belarus\\\small e-mail: sanishchenko@mail.ru, gurinovich@inp.bsu.by}
\date{}

\begin{document}
\maketitle
\begin{abstract}
Design of helical flux compression generator with double-end initiation is described. The developed design approach makes it possible to increase the power in the load by 60\% and shorten the rise time of the pulse by 83\% in comparison with a conventional helical FCG that is made in the same dimensions but with the single-end initiation of explosive. When using an electro-explosive opening switch, the developed FCG allows one to increase the switched current by 35\%.
\end{abstract}

\section{Introduction}
Helical flux compression generators (HFCG), invented by Sakharov in the 50s of the last century~\cite{Sakharov1965,Sakharov1966}, are among the few compact sources of megaampere currents in the world. With the help of HFCG, powerful pulses of electromagnetic radiation are obtained, high-current electron beams are generated, and lightning protection equipment is tested~\cite{Fortov2002,Demidov2012}.

\begin{figure}[ht]
	\begin{center}
		\resizebox{160mm}{!}{\includegraphics{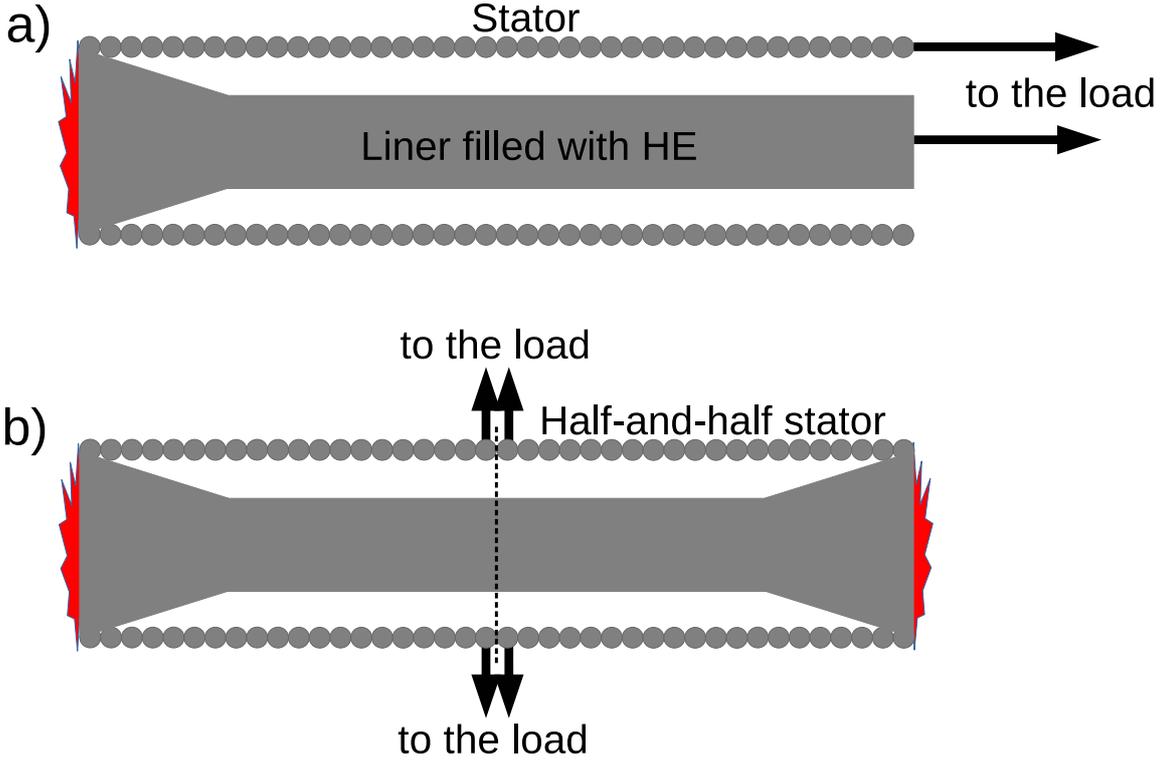}}\\
		\caption{Conventional HFCG with single-end initiation~(a) and HFCG with half-and-half stator and double-end initiation (b).}
		\label{fig:fcg}
	\end{center}
\end{figure}

HFCG comprises a stator (coil) and a metal tube (liner) inserted coaxially inside stator (Fig.~\ref{fig:fcg}a). The liner is filled with high explosive (HE). The operation of the HFCG is as follows. First, an initial magnetic flux is created in the volume between the liner and the stator. Single-end initiation of HE leads to a conical expansion of the liner, which comes in touch with stator, thus locking the flux in the above volume.  Liner expansion is accompanied by compression of magnetic flux. As a consequence, the electric current $I(t)$ and energy $W_l$ delivered to the load increase.

The existence of an upper limit for the detonation velocity imposes limitations on the output characteristics of HFCG. To overcome these limitations, a generator with the double-end initiation of HE was proposed~\cite{Crawford1968,Demidov2010}. The use of double-end initiation makes it possible to double HFCG inductance derivative. As a result, the rate of compression of the magnetic flux increases and the rise time of the current pulse in the load are shortened.

In the present article, a design approach is proposed to compare output produced by HFCG with double-end initiation and that provided by conventional HFCG initiated from one side. Two HFCGs differ by stator only (Fig.~\ref{fig:fcg}). HFCG with double-end initiation has half-and-half stator consisting of two identical parts mirrored to each other. Length of each section of half-and-half stator is half as many as the corresponding section of conventional HFCG stator, which has four winding sections. Length of each section of conventional HFCG is equal to the stator diameter. The total length of both stators is the same.

Using the simulation approach described in \cite{Anishchenko2018}, we will demonstrate that HFCG with double-end initiation makes it possible to generate pulses with faster rise time and higher power delivered to the inductive load than HFCG with single-end initiation, when designed in the same dimensions. In conclusion, it will be shown that HFCG with half-and-half stator and double-end initiation makes it possible to increase significantly the amplitude of the current switched by an electro-explosive opening switch (EEOS).

\section{Conventional HFCG}
Conventional HFCG, which is used for comparison, is a four-section generator with an output section providing matching to the load (hereinafter referred as bucket). The length of each section is equal to the diameter of the stator $D_s=82$~mm. The diameter of the copper liner with 3 mm thick wall is half as many as $D_s$. The liner is filled with an explosive having detonation velocity 7.96 km/s and density 1580~kg/m$^3$.

\begin{figure}[ht]
	\begin{center}
		\resizebox{80mm}{!}{\includegraphics{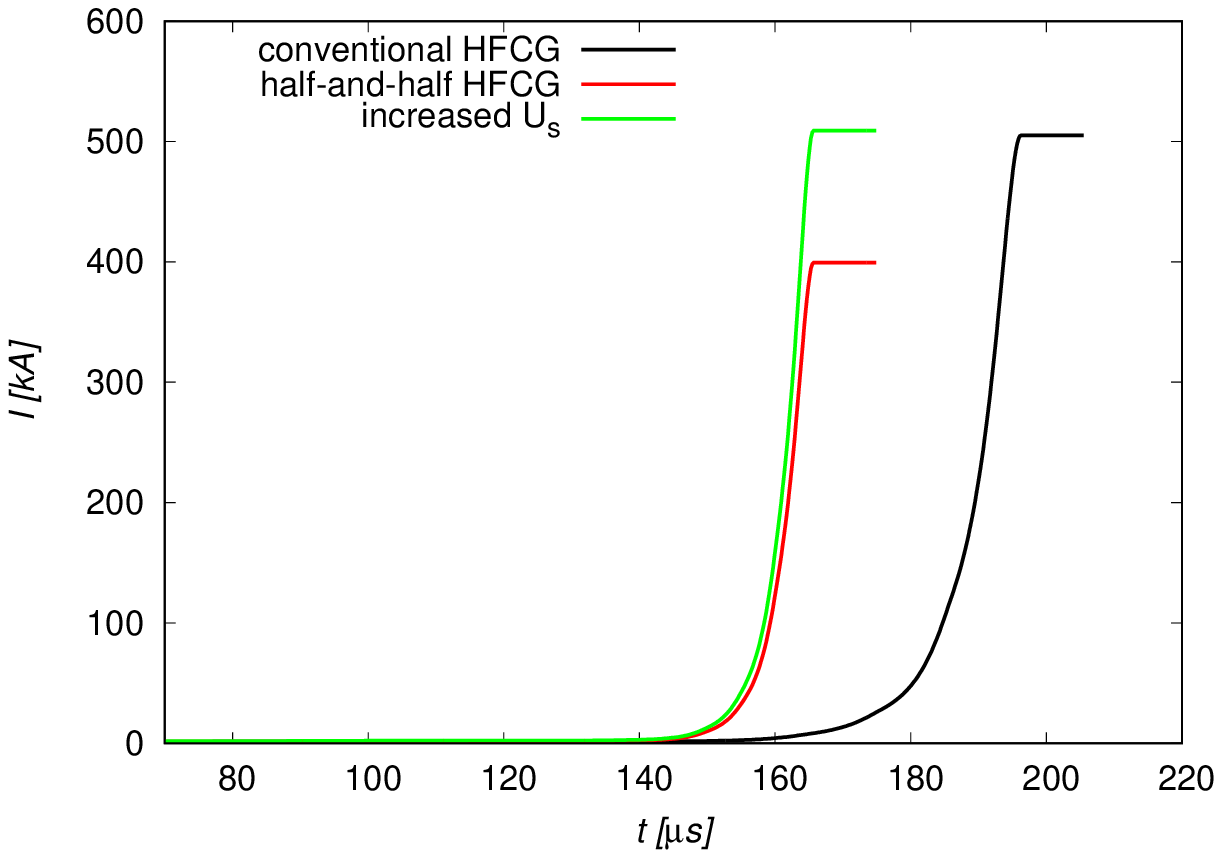}}\resizebox{80mm}{!}{\includegraphics{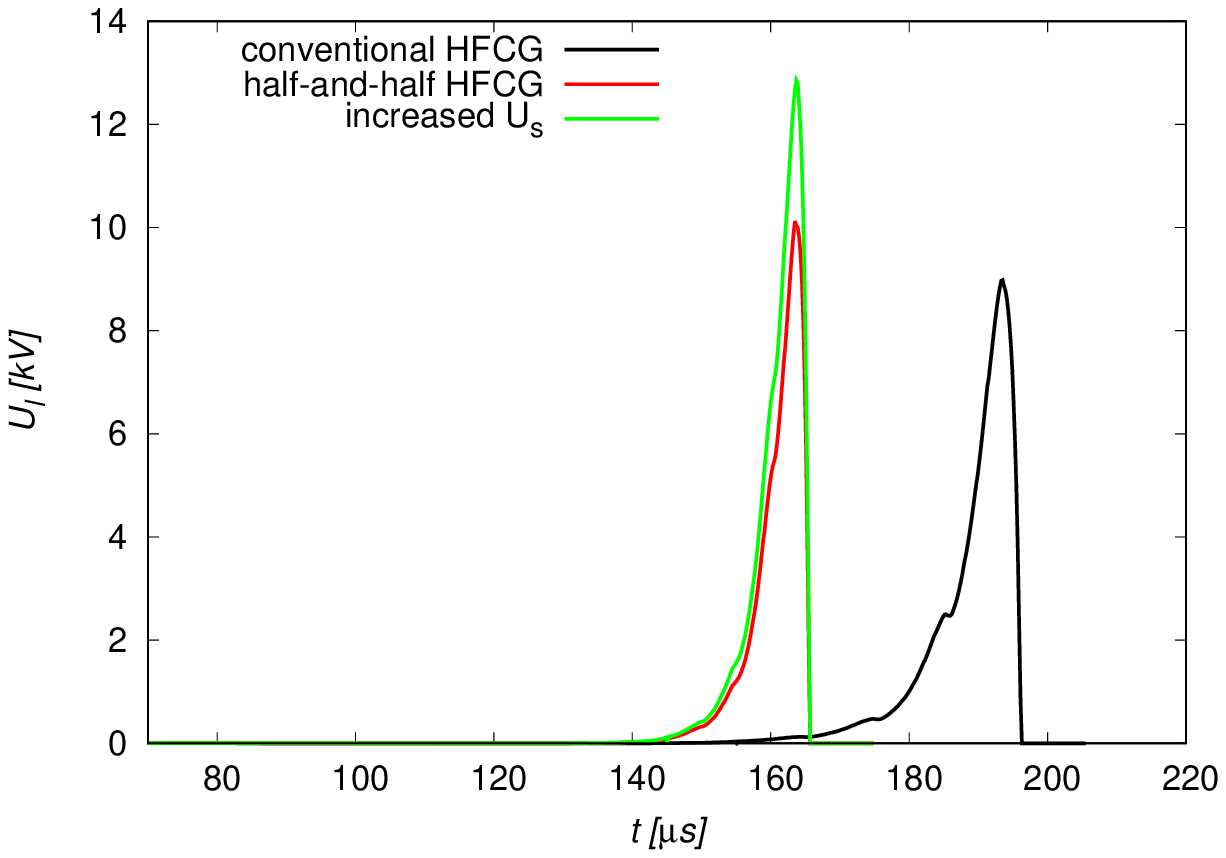}}\\
		\caption{HFCG current (left) and load voltage (right).}
		\label{fig:IgUl}
	\end{center}
\end{figure}

All stator sections are <<wound>> with copper wire. The wire has a metal core diameter 1.4 mm and insulation thickness 0.2 mm. The number of wire starts in the first, second, third, fourth sections and the bucket is 1, 2, 4, 8, 96, respectively.

Seed electric current in the generator, which has initial inductance 157 $\mu$H, is created by discharging a capacitor bank with capacitance 60 $\mu$F. The capacitor bank is charged to voltage $U_s=3$~kV. The moment of explosive initiation is chosen to ensure the electric current at the crowbar closuring instant to have its maximum value. (Crowbar is the area where the liner and stator come into contact for the first time.) Charging voltage is as low as $U_s=3$~kV to ensure the linear current density in HFCG to be lower than 34~MA/m. At higher linear current density values the losses due to the nonlinear diffusion of the magnetic field increase significantly~\cite{Novac2003}.

\begin{figure}[ht]
	\begin{center}
		\resizebox{80mm}{!}{\includegraphics{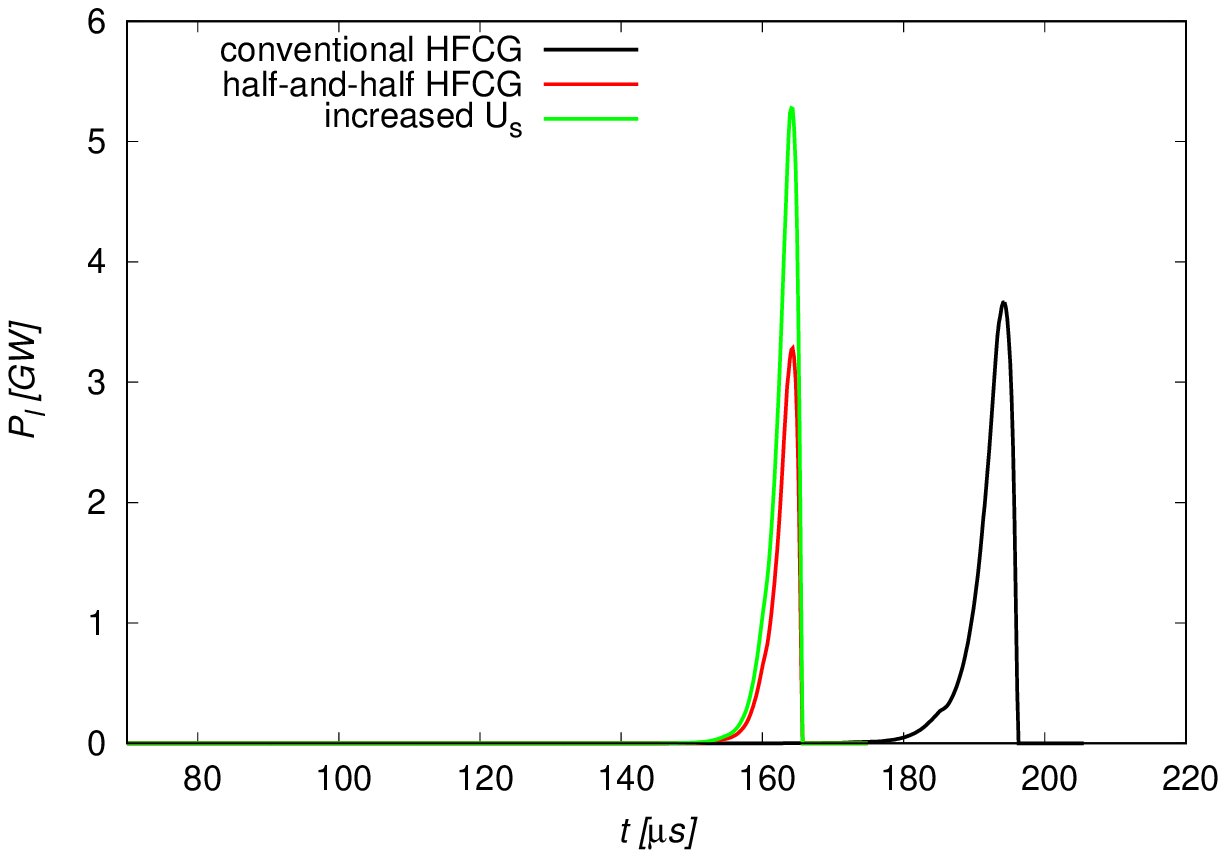}}\resizebox{80mm}{!}{\includegraphics{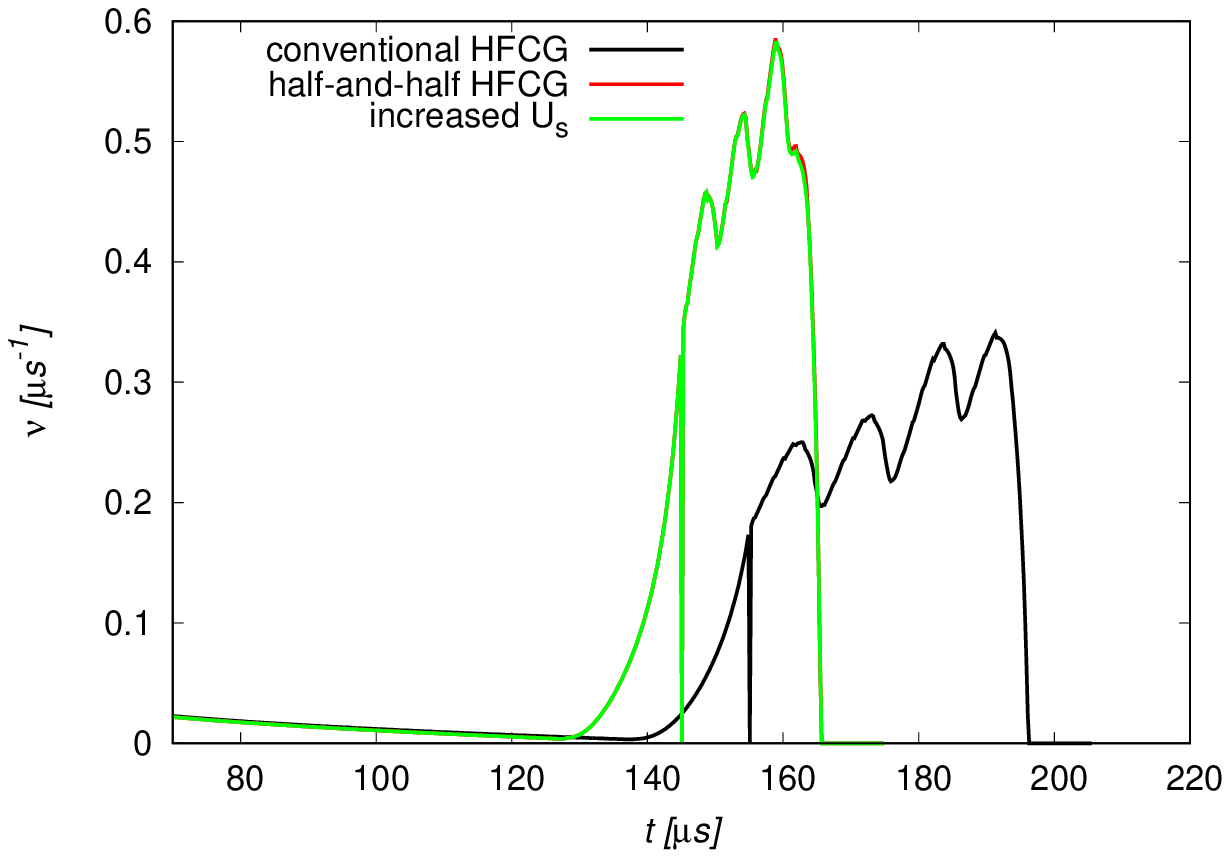}}\\
		\caption{Load power (left) and energy increment (right).}
		\label{fig:PlNu}
	\end{center}
\end{figure}

As a result of current and energy amplification, the maximum load power reaches 3.7~GW (see Fig.~\ref{fig:PlNu}). The load is assumed to be purely inductive. Its inductance is set to be $L_l=150$~nH. In this case, energy increment $\nu=\frac{P_l}{W_l}$, which characterizes the rise time, turns out to be equal to 0.3~$\mu$s$^{-1}$ (see Fig.~\ref{fig:PlNu}).

\section{HFCG with double-end initiation}
HFCG with double-end initiation has the half-and-half stator consisting of two identical parts mirrored to each other, length of each section of half-and-half stator is half as many as the corresponding section of conventional HFCG stator. Initial inductance of HFCG with the half-and-half stator is 13\% lower than the inductance of the conventional generator. At the instant of crowbar closure, this difference reaches 36\%.

Lower initial inductance results in lower current amplification if the load inductance is the same. Therefore, the same charging voltage ($U_s=3$~kV) provides the maximum current in HFCG with double-end initiation to be 20\% lower than in conventional one (see Fig. 2).
Even though the maximum load power is not changed much. At the same time, the rise time is significantly faster: the value $\nu=0.55$~$\mu$s$^{-1}$, which is 83\% higher as compared to conventional HFCG (see Fig.~\ref{fig:PlNu}).

Increase of charging voltage to $U_s=3.9$~kV makes the maximal current value in HFCG with double-end initiation the same as that for conventional HFCG i.e. 500~kA. At the same time, the maximum load power appears increased to 5.3~GW, i.e. by 60\% compared to the conventional HFCG (see Fig.~\ref{fig:PlNu}) with 500~kA maximal current.

\section{Electroexplosive opening switch}
The rise time value is extremely important for operation of EEOS which is an array of thin metal wires or foils connected in parallel. When current with density $j_w>10^{11}$ A/m$^2$. is delivered to array of thin conductors, electric explosion happens after rapid heating~\cite{Fortov2002, Tucker1975}.

According to~\cite{Tucker1975}, electric explosion, which is accompanied by sharp increase in EEOS resistance, occurs if the action integral $A=\int_0^tj_w(\tau)d\tau$ reaches the value $A_{e}=1.24\cdot10^{17}$~A$^2 \cdot$s/m$^2$.
The voltage drop over EEOS is increased, thus, enabling to switch current pulse to a load.

Let us consider the operation of the EEOS when a current pulse is created by HFCG. In the case of exponential increase of electric current in the generator, the current density in thin conductors is approximately described by the relation
\begin{equation} 
j_w=j_ee^{\nu t/2}
\end{equation}
until the moment of electric explosion. (Symbol $j_e$ denotes the current density at the moment of electric explosion.) At the moment of electric explosion, the current density $j_w$ corresponds to the action integral
\begin{equation} 
\label{Ae}
A_e=\int_{-\infty}^0j_e^2e^{\nu \tau}d\tau=\frac{j_e^2}{\nu}.
\end{equation}
Using \eqref{Ae}, we find
\begin{equation}
j_e=\sqrt{\nu A_e}.
\end{equation}

Thus, current density turns out to be proportional to the square root of energy increment $\nu$ at the instant of explosion. Since HFCG with half-and-half stator has $\nu$, which is 83\% higher than that for conventional HFCG, the electric current in EEOS, being proportional to je, is expected to be 35\% greater. Therefore, ceteris paribus, more electric current is switched to the load.

\begin{figure}[ht]
	\begin{center}
		\resizebox{160mm}{!}{\includegraphics{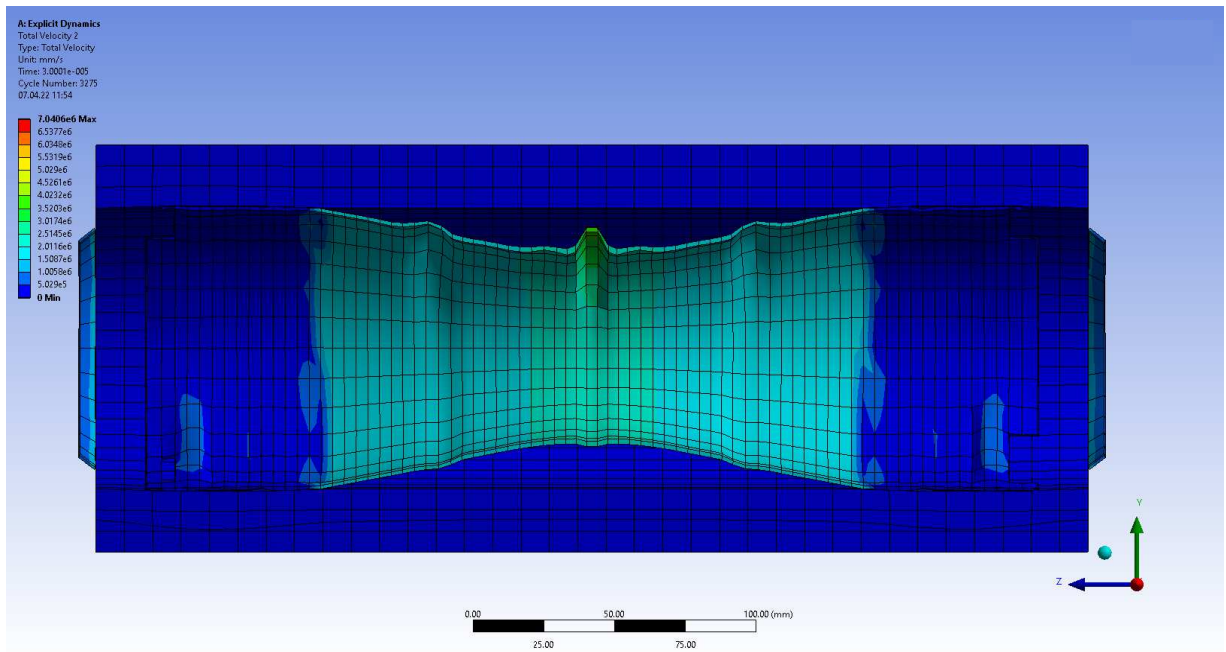}}\\
		\caption{Collision of detonation waves.}
		\label{fig:collision}
	\end{center}
\end{figure}

\section{Collision of detonation waves}

With double-end initiation of HE, two detonation waves propagate towards each other. If we assume that explosive charge fills an entire liner, a collision of two detonation waves is observed in the system. Due to collision, the area with excessive expansion velocity is formed in the liner (Fig.~\ref{fig:collision}). When the liner approaches the stator, the radial expansion velocity is almost 90\% higher than that in the absence of the collision.

It should be noted that in addition to significant increase in the radial expansion velocity liner ruptures are observed during its expansion. Apparently, the ruptures should be accompanied by the formation of jets and inhomogeneities, which both can enhance the probability of electrical breakdown.

In order to avoid degrading the output parameters of the system, the collision of detonation waves should be prevented. This can be achieved by uneven explosive filling of the liner in the area,  where collision of detonation waves is expected.

\section{Conclusion}
A model of HFCG with double-end initiation of HE is described. The proposed design makes it possible to increase the load power by 60\% and shorten the rise time by 83\% as compared to HFCG made with the same dimensions but having single-end initiation of HE. The developed HFCG with double-end initiation allows 35\% increasing of current switched to the load by EEOS.


\begin{thebibliography}{2008}
\bibliographystyle{PhdThesis}
\bibitem{Sakharov1965}
Sakharov A.D. et al., DAN USSR, \textbf{165}(1), 65, 1965.
\bibitem{Sakharov1966}
Sakharov A.D., Sov. Phys., Usp. \textbf{9}, 725, 1966.
\bibitem{Fortov2002}
Fortov V.E., Explosive-Driven Generators of Powerful Electric Current Pulses, (Cambridge International Science Publishing, 2004).
\bibitem{Demidov2012}
Demidov V.A., Plyashkevich L.N., Selemir V.D., Magnetocumulative generators - pulse energy sources. Vol 1. (Sarov: RFNC-VNIIEF, 2012).
\bibitem{Crawford1968}
Crawford J.C., Damerov R.A., J. Appl. Phys., \textbf{39}(11), 5224, 1968.
\bibitem{Demidov2010}
Demidov V.A., IEEE Trans. Plasma Sci., \textbf{38}(8), 1780, 2010.
\bibitem{Anishchenko2018}
Anischenko S.V., Bogdanovich P.T., Gurinovich A.A., Oskin A.V., IEEE Trans. Plasma Sci., \textbf{46}(5), 1859, 2018.
\bibitem{Novac2003}
Novac B.M., Smith I.R., Electromagnetic Phenomena, \textbf{3}(4), 12, 2003.
\bibitem{Tucker1975}
Tucker T.J., Toth R.P., SAND-75-0041. 1975.
\end{thebibliography}
\end{document}